\documentclass[11pt,twoside]{article}

\usepackage{asp2004}
\usepackage{epsf}
\usepackage{psfig}
\usepackage{lscape}
\usepackage{amssymb}

\newcommand{\kms}{\ensuremath{{\rm km\,sec^{-1}}}}                   % km/sec
\newcommand{\msun}{\ensuremath{\mathit{M}_{\odot}}}                  % solar masses: msun
\newcommand{\msunyr}{\ensuremath{\mathit{M}_{\odot}{\rm yr}^{-1}}}   % msun/yr
\newcommand{\rsun}{\ensuremath{\mathit{R}_{\odot}}}                  % solar radius
\newcommand{\mdot}{\ensuremath{\dot{M}}}                             % mass loss rate
\newcommand{\teff}{\ensuremath{\mathit{T}_{\rm eff}}}                % effectieve temperatuur
\newcommand{\vinf}{\ensuremath{v_{\infty}}}                          % maximale uistroomsnelheid
\newcommand{\vesc}{\ensuremath{v_{\rm esc}}}                         % escape velocity

\newcommand{\tdis}{\mbox{$t_{\rm dis}$}}

\newcommand{\be}{\mbox{\begin{equation}}}
\newcommand{\ee}{\mbox{\end{equation}}}

\newcommand{\Mi}{\mbox{$M_i$}}

\newcommand{\Halfa}{\mbox{H$\alpha$}}

\newcommand{\Cref}{\mbox{$m_{\rm ref}$}}

\newcommand{\tfour}{\mbox{$t_4$}}

\begin{document}
\pagestyle{myheadings} \setcounter{equation}{0}
\setcounter{figure}{0} \setcounter{footnote}{0}
\setcounter{section}{0} \setcounter{table}{0}

\title{Mass Loss and Evolution of Stars and Star Clusters:\\
      a Personal Historical Perspective}
\author{Henny J. G. L. M. Lamers}
\affil{Astronomical Institute, Utrecht University, Princetonplein
5, 3584-CC Utrecht, The Netherlands}

\begin{abstract}
The development and progress of the studies of winds and mass loss
from hot stars, from about 1965 up to now, is discussed in a
personal historical perspective.  The present state of knowledge
about stellar winds, based on papers presented at this workshop,
is described. About ten years ago the mechanisms of the winds were
reasonably well understood, the mass loss rates were known, and
the predictions of stellar evolution theory with mass loss agreed
with observations. However, recent studies especially those based
on $FUSE$ observations, have resulted in a significant reduction
of the mass loss rates, that disagrees with predictions from
radiation driven wind models. The situation is discussed and
future studies that can clarify the situation are suggested.
 I also discuss what is known about the dissolution of star clusters in different
 environments. The dissolution time can be derived from the mass and age distributions
of cluster samples. The resulting dissolution times of clusters in the solar neighborhood
(SN) and in interacting galaxies are shorter than predicted by
two-body relaxation of clusters in a tidal field. Encounters with giant molecular clouds
can explain the fate of clusters in the SN and are
the most likely cause of the short lifetime of clusters in interacting galaxies.

\end{abstract}

\section{Introduction}

In this final talk of the workshop I want to take you back about 40 years
and show you the progress of ideas in the two main topics
of this conference: mass loss and evolution of stars and of star clusters.
I will show you that many ideas that are now taken for granted
came as a surprise when the technological progress opened up new possibilities.
I will start at about 1965, when I became involved in astronomical
research
\footnote{From 1962 to 1965 I was the first and only astronomy student at
the University of Nijmegen. Imagine: one professor and one student. The main
interest of the professor was celestial mechanics, not the most interesting topic for an
eager student.
I had the feeling that astronomy could be more fascinating.
So I spent part of my summer vacation of 1965 in the physics library
reading astronomical magazines in search of a topic that would interest me more.
When I read an article by Kippenhahn in ``Sterne und Weltraum'' about stellar evolution
I got so excited that I immediately wanted to switch to that topic. If my astronomy
professor was disappointed that I did not prefer his topic, he did not show it. Instead
he advised me to go to Utrecht University where Anne Underhill had just been appointed
as a specialist in stellar atmospheres and massive stars.}.
(Excellent reviews of the more recent situation have been written by Kudritzki \& Puls, 2000
and by Puls et al., these proceedings, astro-ph/0607290)

\section{Massive Hot Stars: Dull and Not Interesting}

In the 1950s and 1960s massive hot stars did not get much attention. They were rather dull
compared to cool stars and their properties were well understood (or at least
that was the general feeling).\\
-  They did not have chromospheres.\\
-  They were not variable (apart from the Beta Cepheid stars).\\
-  Their optical spectrum showed relatively few spectral lines, mainly of simple ions.\\
-  They all had the same abundances.

Of course, not everything was understood.
There were some puzzling {\it spectral features}:\\
-  Some of the brightest O-stars showed H$\alpha$ in emission, but this was probably
   just a non-LTE effect that was not yet properly understood.\footnote{The study of
   Non-LTE effects in atmospheres of hot stars really started in about 1968 with a
   series of papers by Mihalas and colleagues.}\\
-  Some stars had stronger N and O lines than other stars, but this was probably also
   a non-LTE effect.\\
-  There were some unidentified emission lines, but again these were probably due to
   some non-LTE effect.

Apart from the ``normal'' early type stars, there were also some {\it special
types of hot stars}:\\
-  The {\bf Be-stars} showed emission lines in their optical spectrum, that suggested
   circumstellar disks. These stars were known
   to be fast rotators, so their disks were probably due to the centrifugal force.\\
-  The {\bf Wolf-Rayet stars} with their strong and broad emission lines were already known
   to have a
   stellar wind with a high velocity and high mass loss rate of order $10^{-5}$ \msunyr,
   as shown already in 1934 by Kosirev. However, some astronomers thought that these lines
   were due to a chromosphere and not to a wind.\\
-  The {\bf pathological stars} like $\eta$ Car, P Cygni and the
like were known  to have erupted,
   but the nature of these outbursts and the connection to other stars was unknown.
   They were simply strange exceptions.

In general, there was little interest in spectroscopic studies of early type stars,
apart from their use as tracers of recent star formation.  Most surprisingly,
{\it there was almost nothing known about
the evolutionary connection between these different classes of hot stars!}. For instance,
my teacher Anne Underhill (1966), in her famous book ``The early type stars''
discussed
the observations and properties of all kinds of early type stars but did not mention the
possible evolutionary connections at all!

\section{1967 - 1976: The First UV Observations: All Luminous Early Type Stars
 have Mass Loss!}

 %------------------------------------------------------------------------
\begin{figure}
%\vspace{6cm}
\centerline{\psfig{figure=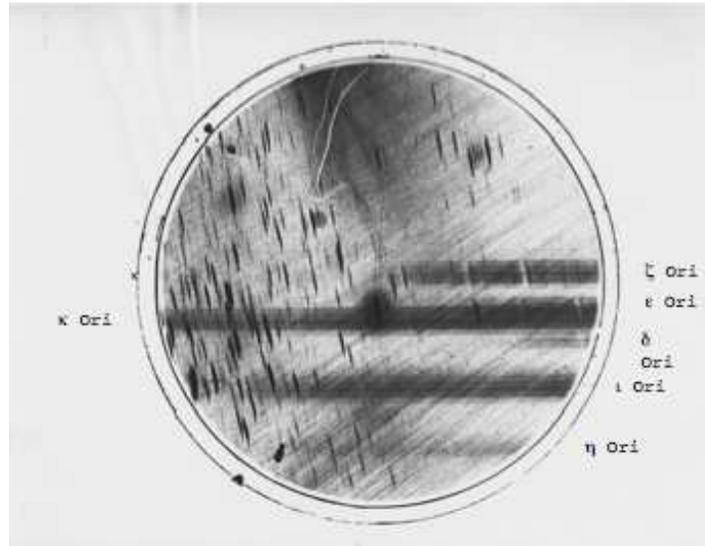,width=9.4cm}}
\caption[]{ The picture that changed our concept of the evolution
of massive stars (Morton 1967a). The horizontal bands are the UV
spectra of five bright stars in Orion: from top to bottom:
$\zeta$,  $\epsilon$ plus $\kappa$ partly overlapping, $\iota$ and
$\eta$ Ori. Wavelength range approximately 1200 to 2000 \AA,
increasing to the right. The P Cygni profiles of Si\,{\sc iv} and
C\,{\sc iv} can easily be seen. The small blotches are the first
order images of Orion stars that were used for wavelength
calibration.} \label{fig:concepta}
\end{figure}
%*****************************************************************************

This picture of rather dull hot stars changed drastically in the
late 60s and early 70s after the first UV spectra were obtained.
Morton (1967a) observed the UV spectra of Orion Belt stars with a
camera in a stabilized nose-cone of an Aerobee rocket over White
Sands. The resulting image (Fig.\ 1) is both awful and
magnificent. It is awful because the flight failed for the second
time and the camera with the photo-cassette  made a hard landing
and was found after two days of searching in the desert. There are
stripes and blotches all over the picture. At the same time, it is
beautiful because it showed for the first time the strong P Cygni
 profiles of UV resonance lines. Using a simple curve of growth analysis Morton (1967b)
estimated mass loss rates of order $10^{-6}$ \msunyr\ with outflow
velocities of 1000 to 3000 \kms. \footnote{I learned about this
discovery in a peculiar way in 1967 when I was a master student of
Underhill in Utrecht, analyzing the optical spectrum of the
supergiant $\epsilon$ Ori. Some Princeton astronomer, called Don
Morton, had come over to talk with  Underhill. She advised him to
talk to me. While walking with him in the park next to the old
Utrecht Observatory, he asked me all kind of questions about the
spectrum of ``my'' star: the abundances, the shape of the spectral
lines, did I notice anything peculiar in the spectrum etc. I was
puzzled and honored at the same time that he had such an interest
in my work. Later in his paper, where he discussed the UV spectra
and the mass loss estimates (Morton et al.\ 1968) he acknowledged
a useful discussion with Mr H.J.\ Lamers.}

The mechanism for the strong stellar winds of hot supergiants was
quickly  identified as radiation pressure. Lucy and Solomon (1970)
were the first to show that the strong UV resonance lines produce
enough radiation pressure to counteract gravity and accelerate the
wind to high velocity by their Doppler-shift\footnote{ In hindsight
this could have been predicted already years earlier by Underhill
and Mihalas (private communications), who both had tried to
calculate hydrostatic  model atmospheres of hot stars but noted
that their program did not converge because the radiation pressure
was too large to allow a stable atmosphere .}.

Within a decade it was clear that mass loss was not limited to the
OB supergiants. Mass loss from A-supergiants was discovered by the
Utrecht UV spectrograph S59, aboard the European TD1a satellite.
The near-UV spectrum of $\alpha$ Cyg showed strong and
blue-shifted resonance lines of Mg\,{\sc ii} with a velocity of
only 200 \kms\ (Lamers 1975)\footnote{I presented this  new result
at a meeting of the Royal Astronomical Society in London and
published it in their rather obscure Philosophical Transactions. I
now advise my students to publish new results first in major
journals and only later in conference proceedings.}.

Then the $Copernicus$ satellite was launched\footnote{I was
fortunate to be a postdoc in Princeton at that time, when the data
of the Copernicus satellite came in.}. Snow and Morton (1976)
published the first catalog of P Cygni profiles which showed that
basically {\it all} early type stars more luminous than $M_{\rm
bol}=-6$ have winds, even the main sequence stars. These
observations also showed that the winds of practically all early
type stars were super-ionized, i.e.\ the degree of ionization was
higher than could be expected on the basis of their effective
temperature (see also Lamers \& Snow, 1978). The high resolution
spectra allowed the first detailed quantitative analysis of the P
Cygni profiles, the first empirical wind model and the first
accurate determination of the mass loss rates of the stars $\zeta$
Pup (O4If) (Lamers \& Morton, 1976) and $\tau$ Sco (B0V) (Lamers
\& Rogerson, 1978). We suggested a simple velocity law, the
$\beta$-law, and found evidence for the presence of strong
``turbulence'' in the winds. We found that the mass loss of
$\zeta$ Pup could not be explained by the observed UV-lines only,
but required the existence of many more lines in the far UV below
912 \AA. This was predicted at about the same time by the
radiation driven wind theory of  Castor et al.\ (1975a).

The observations and the new theory showed that
mass loss would affect the evolution of all massive stars!
That was a very important conclusion that changed the ideas about massive
stars drastically.\\

Within a few years three major steps in understanding the evolution of
 massive stars were taken:\\
- Castor et al.\ (1975b) pointed out that a massive star
throughout its lifetime injects as much energy and mass into the
ISM as a supernova. They also showed that the winds from hot stars
blow bubbles and that
star clusters blow superbubbles in  the interstellar medium.\\
-  Conti (1976) proposed a scenario that linked the different
types of  massive stars into an evolutionary sequence with mass
loss, including the Luminous Blue Variables and the WR-stars:
the ``Conti scenario''.\\
-  de Loore et al.\ (1977) in Brussels and Maeder (1980) in Geneva
calculated the first evolution tracks of massive stars with mass
loss\footnote{Andre Maeder had invited me to Geneva to give a
seminar about mass loss from massive stars. He asked me if it
could be important for stellar evolution.
 Within a year after my visit the first of his famous series of papers
on evolution with mass loss appeared
(but after the first paper on the same topic by the Brussels group).}.\\

\section{1975 - 1980: Stellar Winds studied in Different Wavelength Regions}

Shortly after the discovery that all massive stars have winds
there were  many attempts to quantify the wind parameters, such as
mass loss rate and velocity law. It was realized that this could
be done using observations at different wavelength regions, which
would probe different
regions of the winds.\\

\noindent - Panagia \& Felli (1975) showed that stars with ionized
winds  emit an {\bf excess radio emission}, due to the free-free
process, that could be used to derive the emission measure (EM) of
the wind. Combined with information about the terminal velocity,
derived from spectroscopic UV data, the mass loss rate could be
derived. (This circumvented the difficult problem of the
super-ionization of the stellar winds,
 which plagued the
mass loss studies based on UV lines.)
The radio flux originates far out in the wind where the wind
velocity has reached a constant value.
White \& Becker (1982) later showed in their study of P Cygni
that this model can be tested and the wind temperature
can be derived if the radio image of the wind can be resolved. \\
- Barlow \& Cohen (1977) showed that the winds also produce an
{\bf infrared excess} by free-free emission and derived mass loss
rates from ground-based infrared  observations. This emission is
generated in the lower layers of the wind, where the acceleration
takes place. So its interpretation in terms of mass loss rate
requires an accurate knowledge of the density and velocity
structure in the
lower layers.\\
-  Klein \& Castor (1978) showed that mass loss rates can also be
derived from the equivalent width of the {\bf H$\alpha$ and
He\,{\sc ii} emission lines}. Again this requires knowledge of the
density  and velocity structure in the lower parts of the wind.
This method was later used by Leitherer (1988) and Lamers \&
Leitherer (1993), who adopted the mass loss rates
derived from the radio-flux to calibrate the H$\alpha$ rates.\\
- Cassinelli et al.\ (1978) pointed out that the super-ionization
could be due to {\bf Auger-ionization
by X-rays}. They predicted that hot star winds are  X-ray emitters.\\
- Vaiana et al.\ (1981) detected {\bf X-rays from OB supergiants}
with the {\it Einstein} satellite. The observed X-ray spectra were
interpreted by Cassinelli et al.\ (1981) who showed that the
source of the X-rays is distributed throughout the wind, as
predicted by the shocked wind model of Lucy \& White (1980).

\section{1975 - 1990: Development of Wind Theories}

Already in 1975, at about the time the $Copernicus$ observations
were made, Castor et al.\ (1975a) published their famous theory
that the winds of hot stars can only be explained if they are
driven by a mixture of optically thick and optically thin lines.
This became known as the ``CAK-theory''. It showed that the mass
loss rate could be much higher than the limit of $N\times L/c^2$.
(The limit for mass loss by one optically thick line at the peak
of the Planck curve is about $\mdot \simeq L/c^2$ and the
Copernicus observations showed that there are about $N\simeq 6$
strong wind lines in the UV spectrum at $\lambda > 912$
\AA.)\footnote{ In 1976 I was invited for a seminar at Columbia
University, where Lucy and Solomon (1970) had developed their
model of winds driven by the optically thick winds that were
observed with the rocket experiment by Morton. I mentioned that
the mass loss rate derived for $\zeta$ Pup was much higher than
$6L/c^2$, where 6 is the number of the observed strong lines of
C\,{\sc iv}, N\,{\sc v} and Si\,{\sc iv}. I argued that lines in
the far-UV should contribute significantly to the radiation
pressure. Lucy did not agree and promised ``I will show you within
two months that you are wrong''. I am still waiting.} The CAK
theory was based on the Sobolev approximation and on the
assumption of a ``typical CNO-ion'' for the calculation of the
multitude of optically thick and thin lines. It proved successful
in explaining the trends in observed mass loss rates and wind
velocities.

When the Boulder wind group dissolved\footnote{John Castor  went
to the  Lawrence Livermore Laboratories and Dave Abbott was so
disappointed at the University of Colorado that he decided to
become a primary school teacher.}, the group of Kudritzki and
colleagues in Munich took over the lead in the theories of stellar
winds. They improved the CAK-theory
in two major ways:\\
(1) they dropped the assumption of the star being a point source and took
its finite disk into account (Kudritzki et al.\ 1989),\\
(2) they calculated the strength of an enormous number of lines ($\sim 10^6$)
of many relevant ions (Pauldrach 1987). \\
As a result, their predicted mass loss rates and wind velocities
agreed  much better with observations than the older
CAK-predictions.

Hydrodynamical models of stellar winds by Owocki et al.\ (1988)
improved the original suggestion of Lucy \& White (1980) that line
driven winds are inherently unstable. Fortunately, these
hydrodynamical models also showed that the mass loss rates and
wind velocities predicted by the improved CAK theory were still
correct because they are hardly affected by the presence of
shocks.

\section{1990 - 2000:  Everything fits nicely ! (apart from Some ``Minor'' Problems) }

After the improvements of the observations and wind theories
described above,  the situation seemed
rather satisfactorily in the 1990s: \\
- the basic properties of the winds were known, \\
- the basic mechanism was well understood, \\
- the predictions agreed nicely with the observations,\\
- evolution with mass loss could explain almost all observations.\\
Unfortunately there were two problems that did not seem to be
solved:  super-ionization and clumping.

\subsection{Super-Ionization}

The problem of super-ionization was first raised by the
$Copernicus$  observations which showed strong spectral lines of
high ionization species, such as O\,{\sc vi}, O\,{\sc v}, N\,{\sc
v} and N\,{\sc iv} in the spectra of O-stars and lines of C\,{\sc
iv} and Si\,{\sc iv} in stars down to spectral types B3 (Snow \&
Morton 1976, Lamers \& Snow 1978). These stars are too cold to
create these ions by photo-ionization due to stellar
radiation\footnote{At a conference in Liege in 1978 Jack Rogerson
reported ``The Princeton group had noticed these ions in their
spectra, but we had naively assumed that these could be produced
by the far-UV radiation from the stars. When a young and
unexperienced postdoc looked at the data he immediately pointed
out that this was not possible and that some extra form of heating
was needed''. That postdoc was HJGLML.}.

Originally there were three suggested explanations:\\
- I proposed that the winds of O-stars were ``warm'', with $T \sim
2~10^5$ K, in order to explain O\,{\sc vi} by collisional
ionization in a low density gas and not destroy C\,{\sc iv}
(Lamers \& Morton 1976; Lamers 1979).\\
- Joe Cassinelli suggested that the super-ionization was due to Auger ionization.
He suggested that
hot stars had a thin corona low in the wind (Cassinelli et al.\ 1978). \\
- John Castor suggested  a ``tepid'' wind of $T \sim 6~10^4$ K
that was  optically thick and produced the high ions by
photo-ionization (Castor 1979)\footnote{ There was an interesting
debate at the IAU Symposium 83 at Vancouver Island in 1978, where
the three of us  presented our explanations. We decided to publish
it together, with a score-card showing the pros and contras of
each model (Cassinelli, Castor \& Lamers 1978). It was an exciting
time: three friends
 working closely together with competing models.}.\\
When the X-rays from hot stars were discovered by the $Einstein$
satellite (Vaiana et al.\ 1981), Joe was proclaimed the winner!

However, it soon became clear that the source of the X-rays was
distributed throughout the wind, i.e.\ due to shocks (Cassinelli
et al.\ 1981). This made it difficult to model and explain the
super-ionization because the models of shocked wind were (and
still are) not good enough to predict the ionization fractions
accurately.

The problem became even more severe when the $IUE$ satellite
(1978-1996) observed the spectra of hundreds of early type stars,
but only long-ward of 1215 \AA. This excluded the lines of C\,{\sc
iii}, O\,{\sc vi}, P\,{\sc v}, S\,{\sc vi} and S\,{\sc iv} etc.\
that were observed with the {\it Copernicus} satellite and limited
the mass loss tracers of hot stars effectively to N\,{\sc v},
C\,{\sc iv} and Si\,{\sc iv}. To make things worse, the Si\,{\sc
iv} and C\,{\sc iv} lines are often saturated and provide only a
lower limit to the mass loss rates. The N\,{\sc v} lines are
usually not saturated, but they are from a trace ion that is
sensitive to X-rays of an element whose abundance can change
during the evolution of a star. The determination of the mass loss
rate from these lines
 requires large and uncertain correction factors for its ionization fraction.

The general feeling was that the $FUSE$ satellite, to be launched
in 1999, would solve this problem because it  would observe the
wavelength range down to the Lyman limit where the unsaturated P
Cygni profiles could be observed, just as the $Copernicus$
satellite had done for a small number of stars. Some of these
lines, especially S\,{\sc iv} and S\,{\sc vi} and P\,{\sc v}, are
from trace elements (i.e.\ the lines are not saturated) that are
not affected by changes in the surface composition during the
evolution of the massive stars (but see below).

\subsection{Clumping}

With the mass loss rates derived from UV lines being uncertain,
the attention shifted to the emission lines in the optical
spectrum, mainly \Halfa\ (Klein \& Castor 1978, Leitherer 1988,
Puls et al.\ 1996). However, the detailed analysis of the \Halfa\
profiles soon showed that the strength of the wings of these
emission lines did not agree with the equivalent width (EW) of the
emission (see e.g.\ Hillier 1991; Puls et al.\ these proceedings, astro-ph/0607290).
The EW depends on the emission measure of the wind.  On the other
hand the wings of the emission lines depend on electron column
density.  Adopting a velocity law and using the corresponding
density structure (these are coupled by the equation of mass
continuity) the mass loss rates derived from the wings and from
the EW should give the same mass loss rate. It turned out,
however, that in many (most?) cases they don't.

The mass loss rate derived from the EM is usually larger than that
derived from the wings (Puls et al., these proceedings). This
indicates that the lower layers of the wind, where most of the
\Halfa\ photons are created, is ``clumpy'': the mean value
$<n_e^2>$ is larger than the value of $<n_e>^2$. So obviously, the
structure of the wind is uncertain, especially in the lower
layers,  and the determination of mass loss rates from \Halfa\
profiles is not straightforward.

In principle the radioflux, which is also from free-free emission
and hence depends on $n_e^2$, is also sensitive to clumping.
However, the radioflux comes from far out in the wind and one
might assume that the clumps or shocks due to instabilities deep
in the wind have dissolved by the time the flow  reaches a large
 distance\footnote{This can in principle be checked if the wind can be
 spatially resolved and its
brightness profile can be determined.}.
 So the mass loss rates derived from the radio flux are considered to be the most reliable
ones. Unfortunately the small flux  limited the number of stars
that were observed at radio wavelengths to the brightest ones with
the highest mass loss rates (e.g.\ Abbott et al.\ 1980; Lamers \&
Leitherer 1993; review by Kudritzki \& Puls 2000). With new and
more sensitive radio telescopes this number may increase
drastically.

\section{2000 - now: The state of Confusion}

\subsection{Structures in the Wind?}

The last few days we heard many talks about mass loss rates, which together present
a nice state-of-the-field review. What is my impression? The topic is even more
uncertain than it was before! \\

{\noindent}-  Observations of lines below 1250 \AA\ by the $FUSE$
satellite,  suggest that the mass
    loss rates are ``much'' lower than derived from the ``standard'' UV resonance lines
    by as much as as a factor 3 to 10. This would imply clumping factors of $f \simeq 10 - 100$.\\
-  Part of the problem may be due to the fact that the Sobolev approximation
    is not strictly valid in the complicated winds of OB-stars. For instance, this is a
    basic assumption in the $SEI$ program that is used in several studies
    for calculating and fitting line profiles. The analysis of spectra with more modern
    methods, e.g.\ $FASTWIND$ by Puls et al.\ (2005),  may give more accurate mass loss
    rates (e.g.\ Mokiem et al.\ 2006)\\
-   Another part of the problem may be that clumping might affect the degree of ionization
    of the observed ions.  The X-rays photons that are
    generated in the shocked wind will also affect the ionization.
    An overestimate of the assumed ionization fraction of an ion whose spectral lines have
    been measured, results in an underestimate of the mass loss rate and vice-versa.
     The trace
    ions of dominant elements are expected to be most sensitive to this effect.\\
-  The clumping may be distance dependent. If that is the case, the rates derived from
    \Halfa, from the free-free excess in the IR and the radio regions will all be different.
    There is evidence that this is indeed the case for the star
    $\zeta$ Pup, which is the standard test star for mass loss,
    ever since the first analysis of its $Copernicus$ spectrum.\\
-  Clumping might be  different in different types of OB stars,
e.g.\ the supergiants
    and the main sequence stars. This implies  that even the {\it relative
    mass loss rates} and the trends of mass loss with stellar parameters are uncertain.\\

I wonder how much of this confusion is due to the fact that the winds may be far less
spherically symmetric than is assumed in all studies so far.

Stellar atmosphere models that are used to derive the stellar
parameters (which are input for the wind studies) and the wind
models themselves are always assumed to be spherically symmetric.
Even the most sophisticated wind models with distance dependent
clumping  factors and shocks are still assumed to be spherically
symmetric. What if the wind is much more structured? If that is
the case, the different lines of sight to the star through the
wind might probe different wind structures. For instance, if some
lines of sight to the stellar disk pass through very little wind
material and others pass through the thick wind regions, the UV
line profiles will be weakened by the contribution of continuum
radiation from the lines of sight with low column densities. If
spherical symmetry is assumed in the analysis of such a profile,
the mass loss will be seriously  underestimated.

Is there evidence for non-spherical winds? Certainly! \\
(1) The variable discrete absorption components that are modulated
with the  rotation
period clearly show evidence that the wind has large non-spherical structures.\\
(2) Massive stars may be fast rotating. In this case, not only will the polar region
be hotter than the equatorial regions (due to the von Zeipel effect), but the wind
 from the polar region may also be different from that
of the equatorial regions,  e.g.\ in terms of velocity, density,
shocks, and ionization. In that case the lineprofiles will depend
on the inclination angle to the star, which is usually unknown.

The challenge will be in the next few years to explain the
clumping and confirm or deny the new low mass loss
rates\footnote{I myself am rather skeptical that the mass loss
rates of OB stars are indeed a  factor 3 to 10 smaller than
previously adopted. I think that it would destroy the agreement
between observed and predicted evolutionary aspects of massive
stars including the structure of the bubbles in the ISM.  But
maybe I am just getting more conservative with age?}

\subsection{Mass Loss versus Luminosity}

In the last few years we have seen several papers pointing to the
steep drop in mass loss rate of O-stars in the Magellanic Clouds
at luminosity $\log L / L_{\odot} \lesssim 5 $ (e.g.\ Martins et
al.\ 2004).
 This is usually presented as a  completely unexpected discovery.
The reason that it was unexpected is probably because in recent
years we have started to believe that the mass loss rates scale
with luminosity as a power-law. This was predicted for OB-stars by
the original CAK-theory and by newer predictions of Vink et al.\
(2000). Observed mass loss rates of supergiants and giants
confirmed this trend.

It may be forgotten that the original mass loss observations with
the $Copernicus$ satellite had already shown that, going down
along the main sequence from early-O to late-B, the mass loss rate
suddenly drops by an order of magnitude or more between about
spectral type O9 and B0 (Snow \& Morton 1976). In general,
main-sequence stars later than B0 do not show mass loss signatures
in their UV spectra, unless the star is rotating rapidly (Snow \&
Marlborough 1976). So, there seems  to be a luminosity (?) limit
for high mass loss rates $\mdot \geq 10^{-7}\, \msunyr$.

I wonder if the low mass loss rates of the O-main-sequence stars
in the Magellanic Clouds maybe another manifestation of this same
effect.

\subsection{The Bistability Jump: Does it Exist ?}

Pauldrach \& Puls (1990) noted in their models of P Cygni that the
structure  of the wind changes drastically when they adopted two
slightly different values for luminosity or radius. In one case
the wind was much slower but the mass loss rate much higher than
in the other case. For P Cyg this flip occurs around $\teff\simeq
19300$ K. They called this ``bistability'' because they argued
that the star could jump from one solution to the other and back.
It is due to the drastic change in the degree of ionization and in
the lines that provide the radiation pressure for driving the
wind, mainly metal lines\footnote{I had noticed several years
earlier that the winds of supergiants seem to come in two classes:
with a high terminal velocity, \vinf, of order $10^3$ \kms, or
with much lower \vinf\ of $10^2$ \kms. After the paper by
Pauldrach and Puls on P Cyg I decided to study this in more detail
based on the catalog of P Cygni profiles that we were preparing.}.

Based on this idea, Lamers et al.\ (1995) measured \vinf\ of 68
supergiants in a homogeneous way and calculated the ratio
$\vinf/\vesc$, because that ratio was predicted to depend on
\teff\ in the radiation driven wind models of CAK and the Munich
group. We had to adopt a \teff\ scale based on spectral type. We
found that there was a strong jump in the ratio \vinf/\vesc\
around supergiants of type B1 Ia. Not only the velocity was
drastically different on either side of this type, but more
importantly, so was the observed degree of ionization. The ratio
of the line strength of C\,{\sc ii}/C\,{\sc iii}/C\,{\sc iv}
changed drastically over one spectral subtype, with a high C\,{\sc
ii}/C\,{\sc iv} ratio corresponding to a low value of
$\vinf/\vesc$ and vice-versa. We called it the ``bistability
jump''.

 Vink et al.\ (1999) showed that the jump is
due to the the change in ionization from Fe\,{\sc iv} on the
high-T side to Fe\,{\sc iii}  on the low-T side. Fe\,{\sc iii} has
a much larger number of optically thin lines than Fe IV, which
results in a higher \mdot\ and a lower \vinf. (In terms of the CAK
force multiplier parameters, $k$ increases and $\alpha$
decreases.) When \teff\ of a star decreases due to stellar
evolution and passes the jump temperature, then Fe goes from
Fe\,{\sc iv} to Fe\,{\sc iii}. The resulting higher mass loss rate
and smaller velocity produces an increase in wind density (because
$\rho \sim \mdot/v$) which pushes the ionization even further
down. This is a positive feedback that results in a change in
\mdot\ and \vinf\ in a narrow temperature region of $\Delta \teff
\simeq 2000$ K  ``for any given star''.\footnote{Recently,
radio observations showed the first hint that the bi-stability jump
in terminal velocity is accompanied by a jump in {\it mass loss rate}
(Benaglia, P., Vink, J.S., Marti, J. et al. astro-ph/0703577),
 as predicted by Lamers et al. (1995) and Vink et al. (2000).}

Several groups have improved our study, using larger samples of
stars and, importantly, also using better values of \teff\ (e.g.\
Prinja \& Massa 1998, Crowther et al., these proceedings, astro-ph/0606717). They
find that the jump appears to be less steep then found in our
original study, and that the changes occur over several spectral
subtypes. They conclude that the wind structure changes much less
rapidly with \teff\ than we found. In my opinion, this last
conclusion is due to a misunderstanding of the physical process
that causes the change in the wind structure.

The temperature where this jump occurs depends on the stellar
parameters, e.g.\ the luminosity, mass and radius. This can be
understood easily. A star of higher $L/M$ ratio will have a higher
mass loss rate and hence a higher wind density than a star with
the same \teff\ but a smaller $L/M$ ratio. This means that the
degree of ionization in the first star will be lower and hence the
jump from Fe\,{\sc iii} to Fe\,{\sc iv} will occur at a lower
value of \teff\ (see also Vink et al.\ 2000). The exact value of
\teff\ where the jump occurs will depend on $L$  and $M$ of a
star. So it is no wonder that, as more and more stars of different
$L/M$ ratios are plotted in a diagram of \vinf\ versus \teff, the
jump will become more vague. This is not important. The important
question is: how fast, i.e.\ within how small a \teff\ range, will
the wind change its structure drastically.  The models of Vink et
al.\ (2000) suggest that for each star it will occurs within
$\Delta \teff \simeq 2000$ K.

\subsection{What about the Effect of Mass Loss on Stellar Evolution ?}

When mass loss was discovered, there was excitement and hope that
it  would explain the many unexplained features of hot stars,
e.g.\ the existence of the Humphreys-Davidson luminosity limit,
the appearance of products of the nuclear CNO-cycle at the stellar
surface, the ratio of red to blue supergiants, the existence of
single WR-stars, the trends between numbers of O and WN and WC
stars with galactic distance, etc.

This hope was fully justified. The Geneva group (Maeder, Meynet
and colleagues) published a very impressive series of papers on
the evolution of massive stars with mass loss. They first adopted
in their models the mass loss rates of De Jager et al.\ (1988) but
later the improved rates predicted by Vink et al.\ (2001) were
used, which agreed with the observations of OB stars in the
Galaxy, and the LMC and SMC. Evolution with mass loss could
explain many of the observed features mentioned above. However, it
turned out that mass loss alone could not explain the rapid
appearance of the CNO-products at the stellar surface at the end
of the main sequence phase. It was clear that another effect must
be operating that transports the fusion products to the
surface\footnote{In 1982 when I had redetermined the mass loss
rates of a large number of stars as a function of spectral type
and luminosity class, it was clear that the mass loss rates were
smaller than
 adopted by the Geneva group. I sent a message to Andre Maeder saying
that he should look for an extra mechanism to transport the
nuclear  products to the higher layers, with mass loss doing the
rest of the peeling of the stars. Within a year there was a paper
about evolution with mass loss and convective overshooting, that
could explain the ON-stars.}.

Up to about five years ago massive stars were supposed to rotate
much  slower than critical. This was derived from the broadening
of their spectral lines. However, after Collins \& Truax (1995)
pointed out that the polar regions with their small $v$~sin~$i$
contribute more to the spectrum than the equatorial regions with
their large $v$~sin~$i$, due to the von Zeipel effect, the
rotation speeds were re-evaluated and the O-stars were found to be
closer to critical rotation (see Collins 2004). It was soon clear
that mixing due to differential rotation could explain most of the
features that were originally explained by overshooting (e.g.\
Fliegner et al.\ 1996; Yoon \& Langer 2005; Meynet et al.\ 2006).

Then for a few years almost everything could be explained by the
combined  effects of rotation and mass loss and everybody was
happy again. But now, what if the mass loss rates of OB-stars have
been overestimated by a factor three to ten, as has been suggested
during this conference? Can the agreement between observations and
evolutionary predictions be saved?

There is at least one serious evolutionary problem with the low
mass loss rates. If the radiation driven mass loss rates during
the main sequence phase is so low that the LBV phase is the
dominant phase then it is difficult (or even impossible?) to
explain the strong gradient in the number ratio of WR/O stars with
metallicity from the SMC to the solar neighborhood. Radiation
driven winds will be stronger for higher metallicity stars and
therefore stars in a larger mass range, i.e.\ down to lower
initial masses,  will evolve into WR stars. Therefore the ratio
WR/O stars is expected to {\it increase} with metallicity, if
radiation driven mass loss is important. On the other hand, if
rotation driven mass loss is dominant (e.g.\ during the LBV phase
when the stars eject mass because they reach the $\Gamma
\Omega$-limit due to radiation pressure and rotation) the WR/O
ratio is expected to {\it decrease} with metallicity. This is
because lower metallicity stars rotate faster than higher
metallicity stars (Maeder et al.\ 1999) and so the mass loss would
be stronger for smaller Z. This would produce a dependence of the
WR/O ratio opposite to what is observed!\footnote{This was pointed
out to me by Andre Maeder after the workshop.}

It would be very useful if the evolutionary groups could tell us: \\
- Which evolutionary effect is most critical to the adopted mass loss
 rates of OB stars?\\
- Can this be used to set limits to the mass loss rates?\\
- If the mass loss rates of OB stars are indeed as low as some present suggestions,
  can the observed evolutionary
characteristics  still be explained (e.g.\ compensated by effects due to fast rotation)? \\

\section{ Challenges and Possibilities}

The problems and uncertainties that I mentioned in the previous
sections  imply new challenges for the studies of winds and mass
loss. Here is my personal top list of the challenges and
possibilities:

\begin{itemize}
\item{} Confirm or deny the new reduced mass loss rates. Are they really
a factor 3 to 10 lower than
we have assumed up to now?  If so:\\
- Understand the reason for the discrepancies in the empirical mass
loss rates.\\
- What was wrong with the mass loss rates that were predicted with
the  radiation driven wind models, e.g.\ those derived by
calculating the radiation pressure by following the fate of photon
packages through the wind with Monte Carlo techniques?
\item{} Study the possible effects of a non-spherically structured wind
on the spectral features  (P Cygni profiles, emission lines and
free-free emission) that  are used for deriving mass
 loss rates and compare the results with observations.
\item{} Measure the radio and mm-flux of large numbers of stars of different
 types and classes with
the new instruments. Try to  resolve the sources to study their wind structure.
\item{} Use large spectroscopic surveys to study the mass loss rates in a uniform way. This
will reveal the systematic trends in mass loss and wind
velocities, at least  on a relative scale if not on an absolute
scale, especially if the results can be compared with radio or mm
data.
\item{} Derive the mass loss {\it history} of massive stars by studying the
velocity and density distributions of the circumstellar (CS)
matter around supernovae and GRBs.  Since the wind velocities in
different phases of evolution can differ drastically (e.g.\ $\sim$
2000 \kms\ during the main-sequence phase, $\sim$ 500 to 1000
\kms\ as blue supergiants, $\sim$ 10 to 30 \kms\ as red
supergiant, and $\sim$ 50 to 200 \kms\ in the LBV phase (except
during large eruptions when matter seems to be ejected with a
large range of velocities), the CS matter can reveal the mass
 loss history of the stars (see Vink, these proceedings, astro-ph/0611749).
\item{} If the mass loss rates are indeed lower than has been assumed so far,
 what is the influence
on the evolution of massive stars? Is the LBV phase of massive
stars really  the main mass loss mechanism? Can the observed
properties of massive stars, such as surface abundance, ratios of
O/WR stars etc. be explained with smaller mass rates combined with
fast rotation? (see Sect.\ 7.4).
\item{} Understand the reason for the large radii and the high mass loss rates
of the Wolf-Rayet stars. The near-hydrostatic core of these stars
has a radius $\lesssim$ 1 \rsun. What produces the very  extended
region between this core and the photosphere at $\sim$ 10 to 30
\rsun\ and the resulting high mass loss rate? (see contributions
by Gr\"afener \& Hamann, astro-ph/0609675 and Nugis, these proceedings).
\end{itemize}

\section{And now Something Completely Different: Star Clusters!}

In 1995 I became interested in the evolution of star clusters
while I was on sabbatical at STScI in Baltimore\footnote{I wanted
to use my sabbatical to look for new projects,
 i.e.\ outside the field of
stellar winds. The study of the stellar winds had developed so far
that the interpretation of the observations and the wind models
required a level of complexity that was beyond my ability. I
always liked simple studies based on physical insight.}. I
listened and talked to many colleagues and learned about studies
of extragalactic star clusters with $HST$. When I heard a seminar
about the evolution of Galactic globular clusters, I wondered what
was known about the fate of clusters in other galaxies. Would it
be the same as in our galaxy, even if the conditions are very
different?

A quick study of the literature showed that very little was known about this.
The only studies that I retrieved were those
of Hodge (1986, 1987) and Elson and Fall (1985, 1988) who found that the age distributions of the
clusters in the SMC and LMC are ``wider'' than those of the Galactic open clusters, and
estimated that the decay time of LMC/SMC clusters must be about 5 to 10 times  longer than
those of galactic clusters.

 %------------------------------------------------------------------------
\begin{figure}
%\centerline{\psfig{figure=concept_b3.ps,width=7.0cm},\psfig{figure=conceptgrad_b4.ps,width=7.0cm}}
%\centerline{\psfig{figure=appa_mass_b1.ps,width=7.0cm},\psfig{figure=appa_mass_b2.ps,width=7.0cm}}
%\centerline{\psfig{figure=Appa_age_b1.ps,width=7.0cm},\psfig{figure=appa_age_b2.ps,width=7.0cm}}
\centerline{\psfig{figure=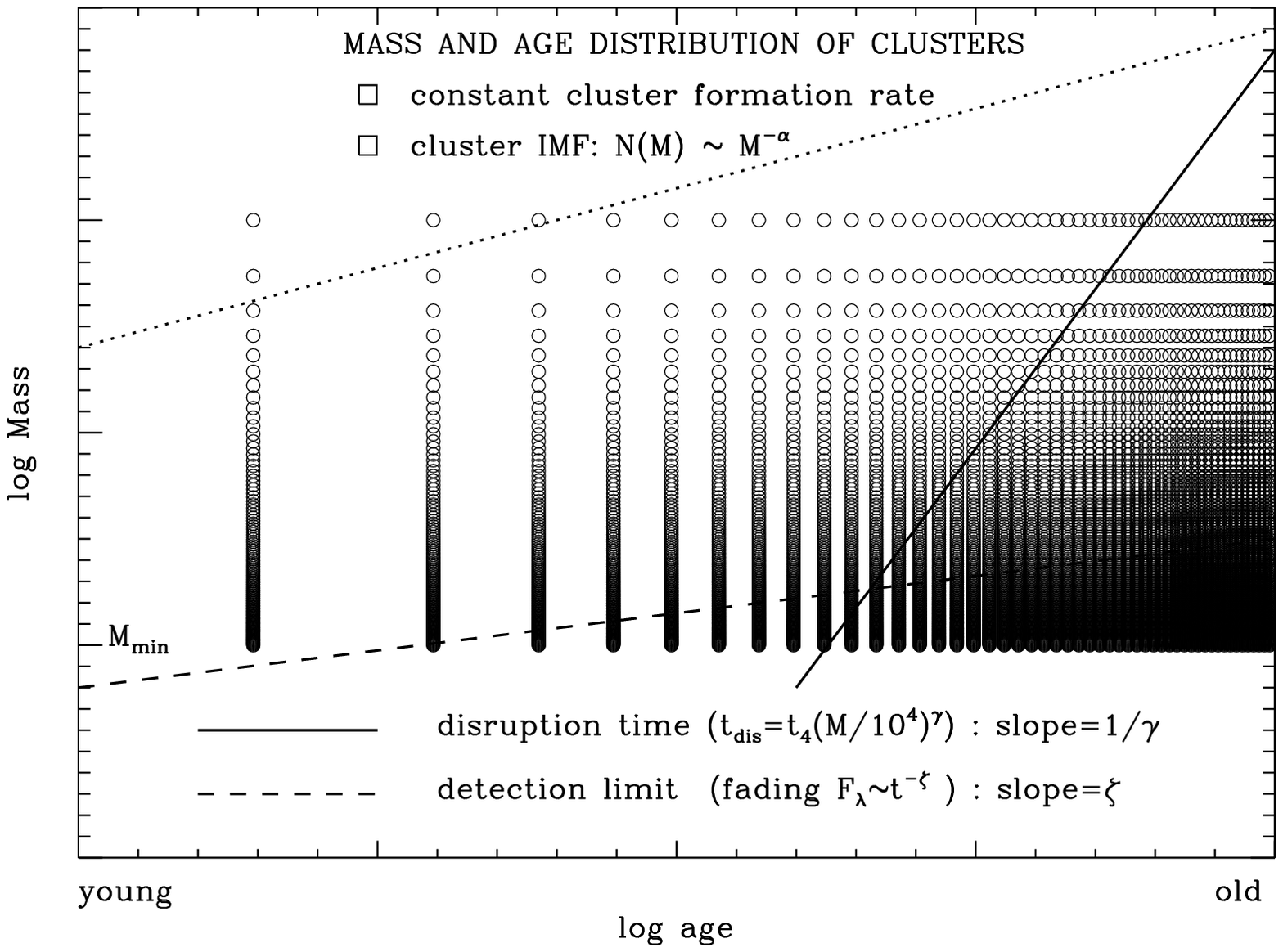,width=7.0cm},\psfig{figure=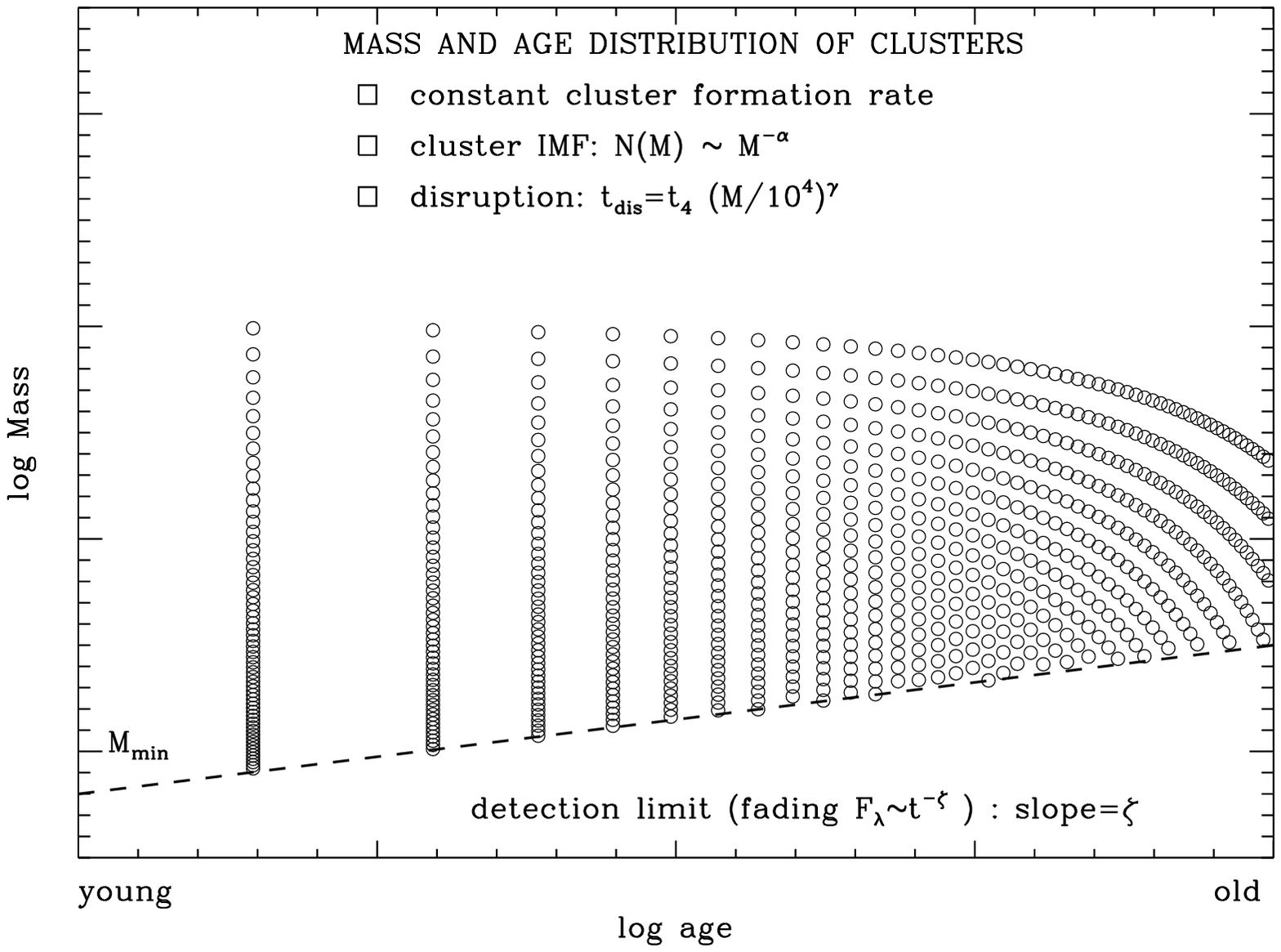,width=7.0cm}}
\centerline{\psfig{figure=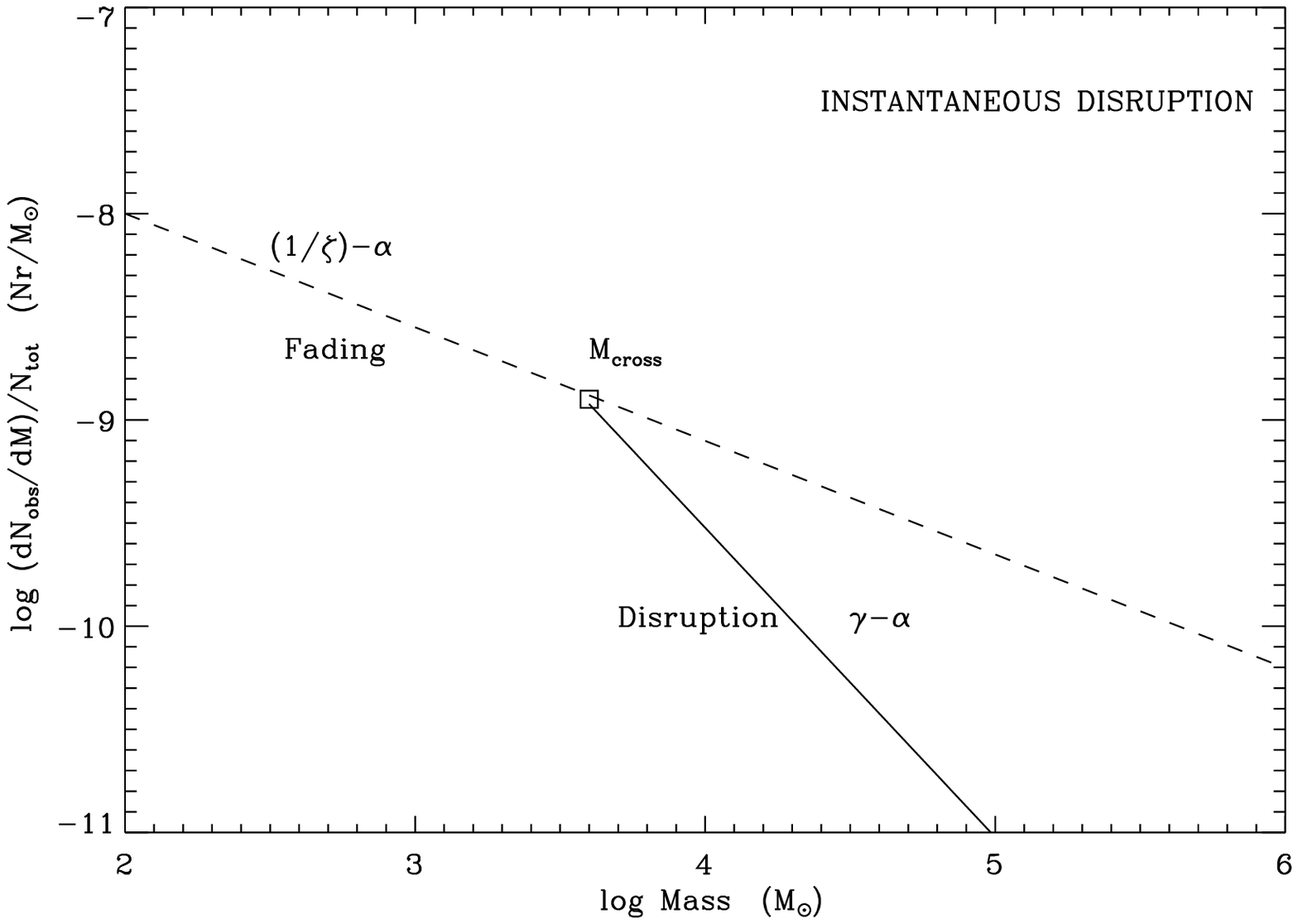,width=7.0cm},\psfig{figure=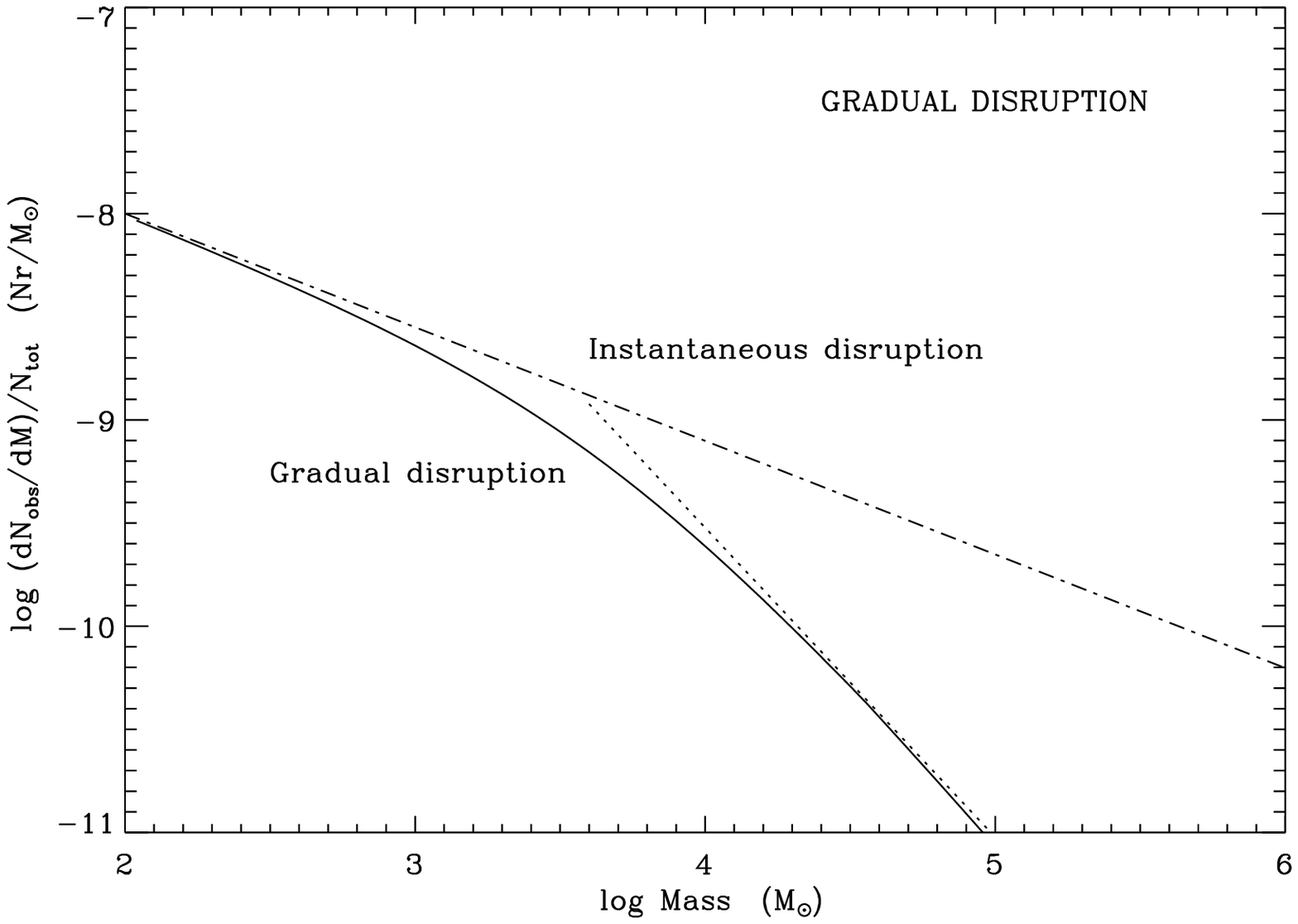,width=7.0cm}}
\centerline{\psfig{figure=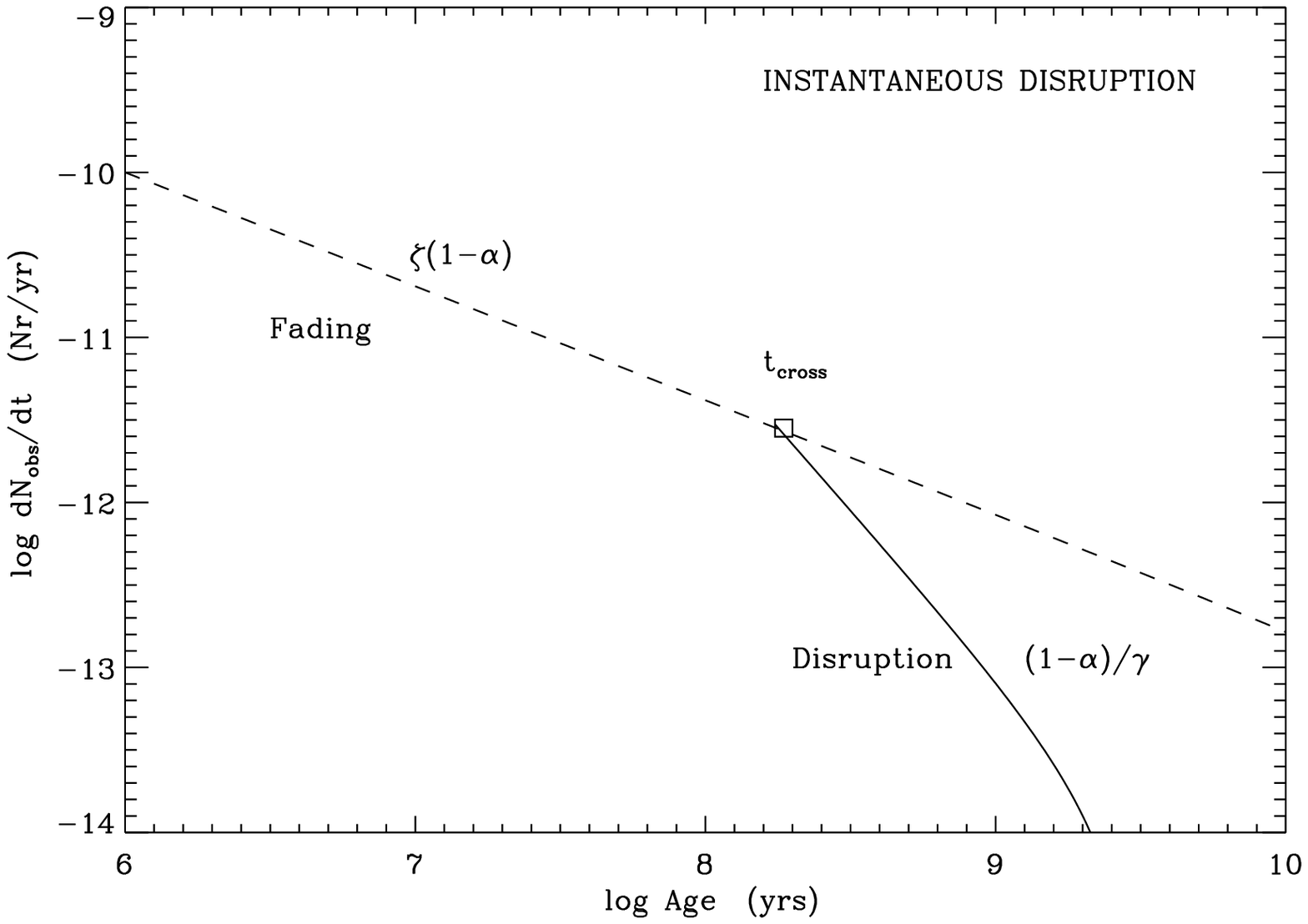,width=7.0cm},\psfig{figure=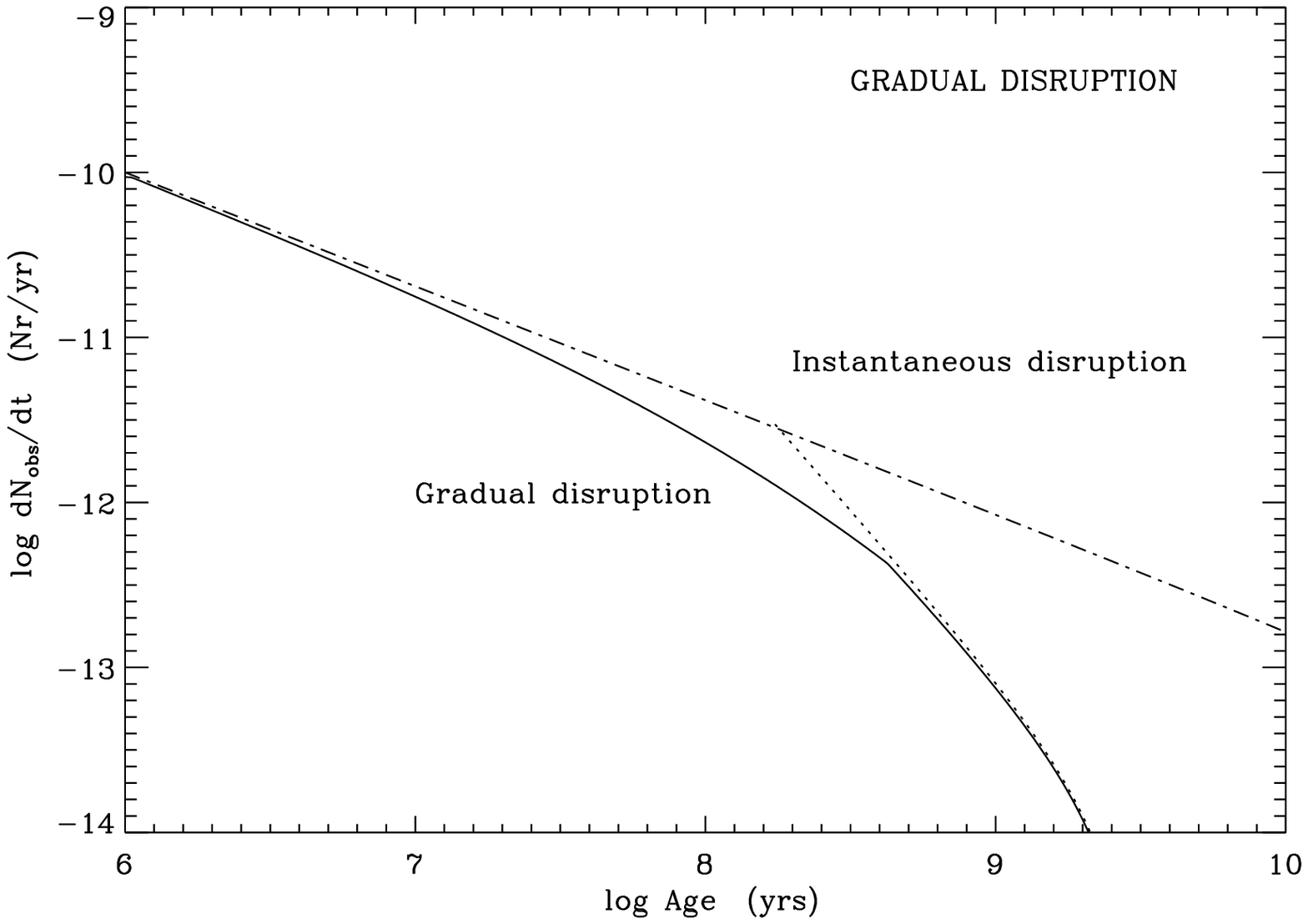,width=7.0cm}}
%\psfig{figure=concept3.ps,width=7.0cm}
%\hfill\vshift{-4.0cm}\psfig{figure=conceptgrad3.ps,width=7.0cm}
\caption[]{ Schematic representation of the Boutloukos \& Lamers (2003) method
for predicting and determining the mass and age distributions of
extragalactic magnitude-limited cluster samples.
Every dot represents a cluster.
{\it Left:} instantaneous dissolution.  {\it Right:} gradual
dissolution with massive clusters dissolving slowly and low mass clusters
dissolving fast.
{\it Top panels:} age-mass distributions.
The upper mass limit
in this diagram will increases with age due to the
size-of-sample effect if the cluster IMF has no upper mass limit.
This is shown schematically in the upper left panel by the dotted line.
{\it Middle panels:} mass distributions.
{\it Lower panels:} age distributions. See text for explanation.}
\label{fig2zk}
\end{figure}
%*****************************************************************************

Back in Utrecht I started to look into the problem with Stratos Boutloukos, a Greek
exchange student. We decided to start in the simplest possible way, in order to
get insight into the dependence of the cluster mass- and age distributions on
the physical conditions. We assumed that:
(a) clusters are formed
 continuously over time with a certain cluster initial mass function (CIMF) of the type
$N(M_{\rm cl}) \sim M_{\rm cl}^{-\alpha}$, and (b) that clusters
have a finite lifetime (dissolution time) that depends on their
{\it initial} mass \Mi\ as a power-law. We chose to normalize this
to the mean value of the cluster masses found in external galaxies
which is about $10^4$ \msun. So $\tdis = \tfour \times
(\Mi/\msun)^{\gamma}$.\footnote{In this first study we adopted that the disruption time depends on the
{\it initial mass} as given by this equation. In the later studies, in which we allowed for
gradual dissolution, we used the same power-law dependence, but now on the {\it present} mass:
 $\tdis = \tfour \times (M(t)/\msun)^{\gamma}$. We also include mass loss by stellar evolution
as $dM/dt = (dM/dt)_{\rm evol}+(dM/dt)_{\rm dis}$ (see e.g.\ Lamers et al. 2005a).}\\
We wondered how the mass and the age distributions {\it of
magnitude limited cluster samples} would evolve over time. In
particular we wanted to know if the values of $\gamma$ and the
constant $\tfour$ could be derived empirically from the observed
distributions of cluster samples of external galaxies?

In order to keep it as simple as possible we started by assuming
a step-function for the evolution of the cluster mass: the mass remains constant
up to the end of its life when the cluster suddenly dissolves. This was of course a
highly simplistic assumption that is physically unrealistic, but it allowed us in
 this first study to gain understanding in the changing age-mass distributions
and its dependence on the CIMF, and the dissolution parameters
\tfour\ and $\gamma$. We adopted the Starburst99 (Leitherer et
al.\ 1999) photometric cluster models to quantify the effects of
fading of clusters due to stellar evolution, until they reach the
detection limit.

The result is systematically shown in the left panel of Fig.\
\ref{fig2zk}. The upper left panel shows the distribution of
dissolving star clusters in  a mass-versus-age diagram for a
magnitude limited cluster sample. Each dot represents a cluster.
The increase in cluster density from high to low mass is due to
the CIMF. The increase from left to right is due to the fact that
the ordinate of the figure is logarithmic in age, so a bin on the
right hand side covers a larger age interval than a bin on the
left side. If the CIMF has no upper mass limit, the observed
upperlimit in this logarithmic age-mass diagram will increase with
age due to the statistical size-of-sample effect: the more
clusters in an agebin, the higher will be the mass of the most
massive cluster. For a CIMF with $\alpha=2$ the maximum mass of a
cluster in an agebin is $M_{\rm max}\propto N$, where $N$ is the
number of clusters in that agebin, so the upperlimit in
logarithmic agebins will increase linearly with age (Hunter et al.
2003, Gieles et al.\ 2006a). This is shown in the top left panel
by the dotted line.

The dashed sloping line represents  the detection limit with a
slope $\zeta$. As clusters get older the evolution of the stars
makes the cluster fainter, with $F_{\lambda} \sim \Mi \times
t^{\zeta}$, with $F_{\lambda}$ proportional to the initial cluster
mass \Mi, and with $\zeta \simeq 0.69$ for the V-band (Leitherer
et al.\ 1999). This implies that clusters can only be detected if
their initial mass was higher than some limit, $\log(\Mi/\msun) >
\zeta \log(t) + {\rm constant}$. Clusters below this limit are too
faint to be detected. The location of this fading line, in terms
of a vertical shift, depends of course on the
known limiting magnitude of the cluster sample.

The full sloping line represents the dissolution time of the
clusters. Clusters of age $t$ have survived dissolution if
$\log(\Mi/\msun) > 4 + \log(t/\tfour)/\gamma$. For a galaxy or a
galactic region where the dissolution time is short, the full line
will be more to the left, whereas it will be located more to the
right for a galaxy with a long dissolution time. Only clusters
above these two limiting lines survived and are bright enough to
be detected. Fortunately, the slopes of the two lines are very
different: the detection limit has a slope of $\zeta \simeq 0.7$,
depending on the wavelength of the limiting magnitude, and the
dissolution line has a slope of $1/\gamma$, which is about 1.6
(see below).

The resulting mass and age distributions can be calculated by
integrating  the distribution in the horizontal direction for each
mass bin  and in the vertical direction for each age bin. They are
   shown in the left middle and lower panels. Because all relations
are power-laws with age or mass, it is easy to see that both
distributions will consist of double power laws, with the kink
being related to the point in age or mass where the two lines in
the top-left panel of Fig.\ 2 cross. The slopes of the double
power laws depend on a combination of the indices  of the CIMF,
$\alpha \simeq 2$, the evolutionary fading $\zeta$ and the
dissolution $\gamma$. With $\alpha$ and $\zeta$ being known, the
values of $\gamma$ and $\tfour$ can be derived from the slopes and
the location of the bend of the empirical age and mass
distributions (Boutloukos \& Lamers 2003).

When we compared this very simple prediction with the age and mass distributions
of observed cluster samples, we found to our surprise that indeed these distributions
showed double power-laws of the type we had predicted! From these distributions
we could derive the dissolution parameters \tfour\ and  $\gamma$ as well as the cluster
formation rates.

The assumption of instantaneous dissolution, adopted in the first
paper, is of course highly unrealistic. It was improved in a
follow-up study, in which we described the decreasing mass and the
fading of a cluster due to both stellar evolution and dissolution
with $dM/dt = (dM/dt)_{\rm evol}+ (dM/dt)_{\rm dis}$ with
$(dM/dt)_{\rm dis} = - M(t)/t_{\rm dis}$ and  $\tdis = \tfour
\times (M(t)/\msun)^{\gamma}$ (Lamers et al.\ 2005a). The
dissolution depends on the {\it present} mass, $M(t)$, of the
cluster, and not on the initial mass $M_i$ as adopted for the
instantaneous disruption model. The result is schematically shown
in the right-hand panels of Fig.\ 2. The mass of all clusters
decreases gradually with age, with the more massive clusters
dissolving slower than the low mass clusters.

The age and mass histograms of these improved models still show
the similar behavior as in the case of instantaneous dissolution,
but the two straight lines that describe fading and dissolution do
not show a kink anymore, but a gradual transition\footnote{The
method of deriving the cluster dissolution together with the
cluster formation history has since been improved  by our group
(see e.g.\ Gieles et al.\ 2005; Bastian \& Gieles, these
proceedings, astro-ph/0609669). We now use the complete density distribution of the
clusters in the mass-age histogram to disentangle the effects of a
variable cluster formation history and cluster dissolution.}.

There were two surprising results of these studies.\\
-  First of all we found that the derived mass dependence of the
dissolution, i.e.\ the exponent $\gamma$, is about the same in
different galaxies, with a mean value of $\gamma=0.62\pm 0.06$. At
about the same time and in the same journal Baumgardt and Makino
(2003) published their results of N-body  simulations of the
evolution
of a grid of clusters in the Milky Way and predicted the same exponent $\gamma=0.62$!\\
- Secondly, even more surprising was the large difference in
dissolution times of clusters in different environments, with
$\tfour$ ranging from 8 Gyr in the SMC to $\sim$ 0.1 Gyr in the
inner regions of the interacting galaxy M51 (Boutloukos \& Lamers
2003; Gieles et al.\ 2005). This was a much wider spread than had
been expected on the basis of two-body relaxations in the tidal
fields of these galaxies (Lamers et al.\ 2005b). Especially the
dissolution time of clusters in the interacting galaxy M51 was
much shorter than predicted.

What could be the reason for this large range in dissolution times
between different galactic environments? Does it mean that
dissolution is dominated by external effects? If so, what are
these effects?

To answer these questions, we studied the age distribution of
clusters in the solar neighborhood, based on the  new catalog of
clusters of Kharchenko et al.\ (2005). We re-derived the
dissolution time of clusters in the solar neighborhood, using an
analytic expression for the mass loss of a cluster due to stellar
evolution and dissolution,  and found  that $\tfour \simeq 1.3\pm
0.5$ Gyr (Lamers et al.\ 2005a). This is much smaller than the
value of 6.9 Gyr predicted by Baumgardt \& Makino (2003) for
dissolution by two body interactions and tidal field stripping,
indicating that other external effects can accelerate the
dissolution of clusters. Could these same effects also be
responsible for the short lifetime of clusters in interacting
galaxies?

Student Mark Gieles decided to study the dissolution of clusters
in different environments by means of N-body
simulations.\footnote{Mark Gieles had the good fortune to be
trained by Lia Athanassoula (Marseille) and Simon Portegies Zwart
(Amsterdam), and he learned very quickly.} He studied the effects
of shocks on the evolution of clusters. This resulted in two nice
(and I think fundamental) papers: one on encounters with giant
molecular clouds (Gieles et al.\ 2006b) and one on shocks due to
the passage through spiral arms (Gieles et al.\ 2007). In these
studies he extended and improved the earlier studies on these
topics by Spitzer (1958), Ostriker et al.\ (1972), Terlevich (1987) and  Theuns
(1991). Most importantly, he showed that a cluster is not
dissolved when the amount of energy, $\Delta E$, added to the
cluster by the shock is equal to
$0.5~E_{\rm pot}$, (as had been assumed before), but that the cluster
is only dissolved if about five times the binding energy is added.
This is because most of the shock energy, about 80\%, goes to
ejected stars with high velocity. When we included the effects of
shocks due to spiral arms and encounters with GMCs in the
predictions of the dissolution time of clusters in the solar
neighborhood, the resulting values of $\gamma \simeq 0.7$ and
$\tfour = 1.7$ Gyr agreed very well with the empirically derived
values (Lamers \& Gieles 2006 and these proceedings, astro-ph/0702166).

These studies have shown that cluster dissolution can be much
faster than predicted by stellar evolution and two body
relaxations only and that the environment plays a crucial role.
This is especially true for violent environments with large
densities of GMCs, e.g.\ in interacting and starburst galaxies!
This has an important consequence. It implies that the
determination of the star formation history of galaxies from the
age distributions of star clusters may lead to wrong results if
the dissolution of clusters is not properly taken into
account\footnote{ Chandar et al.\ (2006a) and Whitmore et al.\
(2007) have recently questioned our results and suggest that they
are due to observational selection effects. Their analysis is
concentrated on ``mass-limited'' cluster samples. However, almost
all empirical cluster samples of distant galaxies, including the
ones we used, are ``magnitude-limited'' and the magnitude limit is
properly taken into account in our studies. See also  the addendum
to Lamers \& Gieles: these proceedings, astro-ph/0702166.}.

It should be realized that the dissolution of star clusters is a
``statistical'' effect. In the same environment some clusters of
the same mass and density may survive longer than others because
encounters with GMCs are random. Therefore the derived dissolution
times have to be considered as ``mean'' values. For instance, the
presence of one or two clusters more massive than expected on the
basis of the mean dissolution time, cannot be used as an argument
that the derived mean dissolution time is incorrect (see e.g.\
Chandar et al.\ 2006b).

All studies mentioned in this section refer to the dissolution of
``bound''  star clusters, i.e.\ clusters that have survived the
infant mortality phase due to the fast removal of gas from the
young cluster.

\subsection{Challenges and Possibilities}

My list of possibilities or challenges for cluster research is
rather short and concerns mainly the studies of cluster statistics
and cluster dissolution. The studies and challenges about the
cluster formation, the shape of the CIMF (power law or
log-normal), infant mortality, early cluster evolution etc. have
been discussed elsewhere in these proceedings by Bastian \&
Gieles  (astro-ph/0609669), Elmegreen (astro-ph/0610687),  
Gieles (astro-ph/0702267), Kroupa (astro-ph/0609370) and Larsen (astro-ph/0609062).

\begin{itemize}

\item{} Study the combined effects of infant mortality and dissolution. Infant mortality
seems to be restricted to ages younger than 10 Myr and is mass
independent. On  the other hand, the dissolution of the surviving
bound clusters, older than about 100 Myr, is clearly mass
dependent. How do clusters in the age range between about 10 and
100 Myr evolve?

\item{} Derive the dissolution times of star clusters in different types of
galaxies and at different locations in the same galaxy, e.g.\ as a
function of galactocentric distance. Compare this with predictions
for different effects of cluster dissolution. This will provide a
check of the dissolution models.

\item{} Study the mass distribution of young cluster samples in a variety of galaxies.
Is it always a power-law of slope $\alpha \sim 2$ or does it depend on the local
conditions? This will not be an easy task, because it requires large samples of
young clusters, automatically restricting these studies to starburst galaxies.

\item{} Study the relation between the age distribution of field stars and clusters.
Some galaxies, e.g.\ LMC, seem to show a different age history for
the
 formation of field stars than for
clusters. This is difficult to explain, because we know that the
vast majority  (if not all) of the stars are formed in clusters
(e.g.\ Lada \& Lada, 2003). Differences in the formation history
of field stars and clusters suggest that the infant mortality rate
may be variable.  For instance, it might be higher or lower during
starburst periods when the star formation efficiency varies with
time.

\item{} Study the photometric evolution of star clusters, taking into account
the fact that dissolution preferably results in the loss of low
mass stars. The resulting photometric evolution may be  different
from that of simple stellar population  models such as Starburst99
or {\it GALEV}.

\end{itemize}

\section{Thanks}

I am grateful to Joe Cassinelli, Mark Gieles and Andre Maeder for
their comments on an early draft of this paper.

I was lucky to have a large number of very good students, in the period
when I studied stellar winds and mass loss, as well as later in my studies of star clusters.
I want to thank them all: ``I learned a lot more from you than you did from me!''

Over the years I collaborated with many colleagues on a variety of
topics. It was nice to see so many of them attending and
contributing to this meeting.  Thank you all for the good times we
had in sharing the excitement of our research. I hope it is not
over yet. I particularly want to mention my friend and co-author
of the book on stellar winds, Joe Cassinelli, who could not be
here.

Last but not least I want to thank the organizers of this nice workshop, especially
the co-chairs of the SOC, Linda Smith and Rens Waters, and the chair and
secretary of the LOC, Alex de Koter and Marion Wijburg.

\end{document}